\documentclass[12pt,preprint]{aastex}

\begin{document}

\title {LARGE CIRCUMBINARY DUST GRAINS AROUND EVOLVED GIANTS?}

\author{M. Jura, R. A. Webb}
\affil{Department of Physics and Astronomy, University of California,
    Los Angeles CA 90095-1562; jura@clotho.astro.ucla.edu; webbr@astro.ucla.edu}

\author{C. Kahane}
\affil{Observatoire de Grenoble, B.P. 53, F-38041 Grenoble Cedex 9, France;
kahane@obs.ujf-grenoble.fr}

\begin{abstract}
 We have detected continuum emission at 450 ${\mu}$m, 850 ${\mu}$m and 1.35 mm from SS Lep (or 17 Lep), 3 Pup and probably BM Gem,  likely or confirmed binary systems consisting  of at least one evolved  giant.  The observed submillimeter fluxes are probably emitted by grains rather than ionized gas.  The fluxes are larger than those expected from a ``normal" dusty wind; the dust temperature is ${\leq}$ 70 K within 6${\arcsec}$ of the stars.  To explain why grains are so cold near the star, we suggest that the  emission at ${\lambda}$ ${\geq}$ 450 ${\mu}$m is  produced
by particles as large as 0.1 mm in radius, and that these large particles probably have  grown by coagulation in
circumbinary orbiting disks with masses ${\geq}$ 5 ${\times}$ 10$^{28}$ g.      
\end{abstract}


\section{INTRODUCTION}

The growth of solids into planetesimals in circumstellar disks   is a major unsolved astrophysical problem.  Almost all investigations of dust disks have concentrated on pre-main sequence stars, but there are some post-main sequence binary stars which possess
orbiting circumbinary dust disks.  We are studying 
 disks in evolved systems with the goal of learning more about
particle growth.

Accretion disks around a mass-receiving star in mass-exchange binary systems such as cataclysmic variables and
symbiotics stars have been studied for many years.   In an accretion disk, the lifetime of an individual grain may be short even if the system is long-lived.  In contrast,  in a circumbinary disk, the particles orbit both stars and are not  being continuously destroyed.  As a result, in circumbinary disks, particles
may survive enough orbits to grow by coagulation to sizes as large as 1 mm. With only a few known examples of this phenomenon such as the Red Rectangle and AC Her 
(see Waters et al. 1993, Jura \& Kahane 1999, Jura, Chen \& Werner 2000, Molster et al. 2000),  we hope to identify more such systems.

Because  small spherical grains do not emit efficiently at wavelengths much larger than their size (see, for example, Spitzer 1978), one way to identify ``big" particles is to observe at relatively
low frequencies.  
 Therefore,  
we have obtained mm and sub-mm observations of SS Lep (m$_{V}$ = 5.0 mag, period = 260 days, A1 + M4; Cowley 1967, Welty \& Wade 1995) and 3 Pup (m$_{V}$ = 4.0 mag, period = 161 days, A2I + ?; Plets et al. 1995) because they are  A-type stars in the Yale Bright Star Catalog with  anomalously and uniquely  high IRAS fluxes  (Jura \& Kleinmann 1990) and we suspected that they may have circumbinary disks.  As part of this program, we also obtained data for BM Gem, a highly luminous carbon star with oxygen-rich circumstellar matter which also may  possess a long-lived orbiting disk (Kahane et al. 1998).          
  
There are many previous infrared studies of evolved binary stars (for example, Friedemann, Gurtler \& Lowe 1996).
A system that might have some properties in common with SS Lep and 3 Pup is the binary  containing the post-main sequence luminous A-type star, ${\epsilon}$ 
Aur (M$_{v}$ ${\leq}$ -6.0 mag) which has a companion of unknown type that is surrounded by a dust disk containing grains larger than 5 ${\mu}$m (Lissauer et al. 1996).  Here, we consider circumbinary environments where the grains may be more than an order of magnitude larger in size and thus 10$^{3}$ times more massive than those found around the companion to ${\epsilon}$ Aur.     
 
In contrast to their infrared properties, little is known about submillimeter dust emission from evolved binary stars.
Symbiotic stars have been detected at ${\lambda}$ $>$ 100 ${\mu}$m (Seaquist \& Taylor 1992, Ivison et al. 1995, Corradi et al. 1999), but this emission is probably produced by ionized
gas and not dust.  In the systems discussed here, thermal emission by dust probably dominates at ${\lambda}$ ${\leq}$ 1350 ${\mu}$m.   

\section{OBSERVATIONS}

The 1.35 mm flux measurements were obtained on  1999 Jan 21 in the service observing mode with the SCUBA  camera (Holland et al. 1999) using the photometric mode on the  James Clerk Maxwell Telescope, or JCMT, at Mauna Kea, Hawaii.  
  The telescope half-power beam diameter at 1.35 mm was 19{\tt"}. The atmospheric opacity was scaled from simultaneous measurements from the Caltech Submillimeter Observatory radiometer and ranged between 0.11 and 0.13. The
data were reduced and calibrated using the ``SCUBA User Reduction Facility" (or SURF, Jenness 1998).  The  secondary calibrator, RAFGL 618, was used to calibrates fluxes with an assumed flux of 2.49 Jy.  Integration times of 30 and 36 minutes yielded the measured fluxes reported in Table 1. 

The 450 ${\mu}$m and 850 ${\mu}$m fluxes were measured on  2000 Jan 31, again
at the JCMT in the service observing mode with the SCUBA camera in the photometric mode following procedures similar to those for our previous observations.  The  secondary calibrators were RAFGL 618 and OH 231.8+4.2 with assumed fluxes of 4.57 Jy
and 11.9 Jy at 850 ${\mu}$m, respectively and 2.52 Jy and 10.5 Jy at 450 ${\mu}$m. 
The half-power beam diameters were 8{\arcsec} and 12{\arcsec} at 450 ${\mu}$m and 850 ${\mu}$m. The results are reported in Table 1. The listed uncertainties represent the RMS noise levels, and do not include an estimated 10\% calibration error at 1.35 mm and 850 ${\mu}$m and a 20\% calibration error at 450 ${\mu}$m.  With an extragalactic background at 850 ${\mu}$m of probably fewer than 5 sources degree$^{-2}$ brighter than 50 mJy (Smail, Ivison \& Blain 1997), it is unlikely that source confusion contaminates our observations. 

\section{STELLAR PARAMETERS}

For many years it was thought that the A-type star in  SS Lep lies on the main sequence; it has a luminosity class V in the Yale Bright Star Catalog.  However,  the distance 
measured with the {\em Hipparcos} satellite is 330 pc.    
With A$_{V}$ ${\sim}$0.4  mag (see Blondel, Talavera \& Tjin A Djie 1993) and m$_{V}$ = 5.0 mag, then $M_{V}$ =  -3.0 mag and thus this A-type star lies well
above the main sequence  (see Perryman et al. 1995). 
Analysis of the Doppler motions show that the A-type star has 3.5 times the mass of the M-type companion (Welty \& Wade 1995), and it is unlikely that both stars are in pre-main sequence evolution since the time scales for remaining above the main sequence are such a strong function of stellar mass (Palla \& Stahler 1993).  Instead, it appears that
this is an interacting binary where both stars have evolved beyond the main sequence (Pols et al. 1991).  The 
binary companion to 3 Pup is unseen (Plets et al. 1995), while
it is only suspected that BM Gem possesses a companion (Kahane et al. 1998).  Our estimates for the distance, $D_{*}$, effective temperature, $T_{*}$,
and radius, $R_{*}$, of each star along with relevant references are given in Table 1.

\section{THE EVIDENCE FOR LARGE GRAINS}

 The spectral energy distribution of SS Lep is shown in   
Figure 1; the plots for BM Gem and 3 Pup are similar.     For the sake of comparison, we also display in this Figure the scaled spectral energy distributions for VY CMa and the Red Rectangle, as representative of  the emission from a ``normal" wind (where the density varies as $R^{-2}$ where $R$ is the distance of the grains from the star) and  a system with large, orbiting grains, respectively. In Figure 1, we ignore the 2.2 ${\mu}$m variability of SS Lep which is at most a factor of 1.4 (Kamath \& Ashok 1999, Neugebauer \& Leighton 1969, Malfait, Bogaert \& Waelkens 1998).

The fluxes at ${\lambda}$ $>$ 100 ${\mu}$m  are probably the result of grain emission rather than ionized gas. At ${\lambda}$ ${\leq}$ 1.3 mm for SS Lep,  $F_{\nu}$ increases 
 as rapidly as  ${\nu}^{+1.6}$.  Thus, this continuum cannot arise from optically thin ionized gas since F$_{\nu}$ would be approximately flat.  It is also unlikely that the continuum is produced by optically thick ionized gas with optical depth, ${\tau}$.
With a Gaunt factor of 2.5 (Karzas \& Latter 1961), to account for the observed value of F$_{\nu}$(450 ${\mu}$m) by  free-free emission from a gas of ionized hydrogen at 7500 K,  the required value of $n_{e}^{2}V$, where $n_{e}$ is the electron density in volume $V$, is 9.6 ${\times}$ 10$^{57}$  ${\tau}/(1 - e^{-{\tau}})$ cm$^{-3}$ .  With ${\tau}$(450 ${\mu}$m) ${\geq}$ 1.2 to account for the observation that F$_{\nu}$(450 ${\mu}$m)/F$_{\nu}$(850 ${\mu}$m) ${\approx}$ 2.5, then   
$n_{e}^{2}V$ ${\geq}$ 1.7 ${\times}$ 10$^{58}$ cm$^{-3}$.  The H${\alpha}$ emissivity of a gas at 7500 K is 4.7 ${\times}$ 10$^{-25}$ erg cm$^{-3}$ s$^{-1}$ (Osterbrock 1974) so the predicted total H${\alpha}$ luminosity is 7.8 ${\times}$ 10$^{33}$ erg s$^{-1}$.  With an  H${\alpha}$ emission equivalent width of ${\sim}$ 3 {\AA} (see Figure 1 in Welty \& Wade 1995) and the stellar parameters in Table 1, the actual H${\alpha}$ luminosity of SS Lep is 2.6 ${\times}$ 10$^{33}$ erg s$^{-1}$, about a factor of 3 less than predicted.

While the observed  emission at ${\lambda}$ ${\leq}$ 1.35 mm is probably produced by dust, the measured fluxes at ${\lambda}$ ${\geq}$ 450 ${\mu}$m  are too strong to be produced from a wind similar to that around VY CMa. Detached shells which are much more extended  than the JCMT beam such as those proposed by van der Veen et al. (1994) could conceivably explain the presence of a large amount of cold dust.  However, as explained in more detail in Section 5, such detached shell models predict more neutral gas than is observed.             

Given that neither a ``normal" wind nor ionized gas can easily explain our measurements, we present an alternative.    Assume, for simplicity, that the grain emissivity as a function of frequency can be described as a power law, ${\nu}^{p}$.  Equivalently, if $Q_{\nu}$ denotes the ratio of absorption to geometric cross section, then $Q_{\nu}$ varies as ${\nu}^{p}$.  If the grains are small compared to the wavelength of the light which they emit, then $p$ ${\sim}$ 1, while larger grains might have $p$ ${\sim}$ 0.  If  $T_{gr}$ denotes the grain temperature at  distance, $R$, from the central  binary, in a steady state where the grain heating, which is determined by the sum of the mean intensity, $J_{\nu}$, from both stars, is balanced by
radiative cooling, then:
\begin{equation}
\int\;Q_{\nu}\;J_{\nu}(R)\,d{\nu}\;=\;\int\;Q_{\nu}\;B_{\nu}(T_{gr})\;d{\nu}
\end{equation} Therefore, summing over the contribution to $J_{\nu}$ by the two stars: 
\begin{equation}
T_{gr}^{(4\,+\,p)}\;=\;(1/4)\;\;{\sum}_{i}\;\;T_{*,i}^{(4\,+\,p)}\,(R_{*,i}/R)^{2}
\end{equation}
where $T_{*,i}$ and $R_{*,i}$ are the effective temperature and radius of
the illuminating stars in the central binary. 
   
To compare the predictions of this simple model,  we expect 
for optically thin observations at two different
frequencies that:
\begin{equation}
F_{{\nu}_{2}}/F_{{\nu}_{1}}\;=\;({\nu}_{2}/{\nu}_{1})^{p}\;(B_{{\nu}_{2}}[T_{gr}]/B_{{\nu}_{1}}[T_{gr}])
\end{equation}
We take F$_{\nu}$(100 ${\mu}$m) as measured in the  large IRAS beam  and compare with 
F$_{\nu}$(850 ${\mu}$m) as measured within the 12${\arcsec}$ diameter JCMT beam to derive an upper bound for the grain temperature at 6${\arcsec}$ from the star as expected from equation (3) for both $p$ = 0 and $p$ = 1.   Especially for 3 Pup and SS Lep, if $p$ = 1, the predicted value of $T_{gr}$ of ${\sim}$ 120 K is significantly larger than the inferred value of ${\leq}$ 45 K. That is, ``small" particles, by themselves, with $p$ = 1 cannot be cold enough within the beam
of the JCMT to explain the relatively strong submillimeter fluxes.  
  In contrast, models with at least some particles with $p$ = 0 can fit the data.

If the grains are spheres of radius $a$, then the requirement that $p$ = 0  implies that $a$ $>$ ${\lambda}$/(2${\pi})$.   For ${\lambda}$ = 850 ${\mu}$m,   then $a$ $>$ 0.1 mm. 
If the grains are orbiting the central binary, then if the disk is optically thin, in order for the gravitational attraction to be larger than the outward force of radiation pressure,
\begin{equation}
a\;>\;3L_{*}/(16{\pi}GM_{*}c{\rho}_{s})
\end{equation}
where $M_{*}$ is the summed mass of the binary and ${\rho}_{s}$ is the matter density of the grains which we take equal to 3 g cm$^{-3}$.  For SS Lep, with $M_{*}$ ${\sim}$ 3 M$_{\odot}$ (see Pols et al. 1991), then $a$ $>$ 0.2 mm.  
 
\section{DUST DISK PARAMETERS}

A natural explanation for the presence of large amounts of cold, large grains is that the particles reside in long-lived  disks.  
Measurements of the submillimeter fluxes allow estimates for the disk parameters. 
For SS Lep, with F$_{\nu}$(1350 ${\mu}$m) = 25 mJy, in order to have a brightness temperature less than 1000 K, the approximate sublimation temperature of the grains, the disk radius inner radius must be larger than 7 ${\times}$ 10$^{13}$ cm.   This size is greater than   the binary separation of 1.5 ${\times}$ 10$^{13}$ cm if the orbital inclination is 33$^{\circ}$ (Welty \& Wade 1995);  the particles must be circumbinary.  The inner disk temperature of 450 K derived from the IRAS data (Fajardo-Acosta \& Knacke 1995)
 implies that  the inner radius of the disk, $R_{in}$, is  3 ${\times}$ 10$^{14}$ cm if $p$ = 0.  

The mass of the grains, $M_{dust}$, that are emitting at wavelength ${\lambda}$ can be estimated by using the Rayleigh-Jeans approximation that:
\begin{equation}
M_{dust}\;=\;(F_{\nu}\,{\lambda}^{2}\,D_{*}^{2})/(2\,k\,T_{gr}\,{\chi}_{\nu})
\end{equation}
where $D_{*}$ is the distance from the Earth to the star and ${\chi}_{\nu}$ is the opacity (cm$^{2}$ g$^{-1}$) of the material.  We adopt ${\chi}_{\nu}$ (850 ${\mu}$m) ${\sim}$ 3 cm$^{2}$ g$^{-1}$ (Pollack et al. 1994).  Assuming an average value of the grain temperature in the telescope beam of 100 K, the mass in dust, $M_{dust}$, is given for each star in Table 1.  In every case,  $M_{dust}$ ${\geq}$ 5 ${\times}$ 10$^{28}$ g.  

In a ``detached shell" model similar to that proposed by van der Veen et al. (1994), we assume that $p$ = 1, and    in Table 1 we list the
value of the radius of a detached dust shell, $R_{det}$, required for the grains to be as cold as the temperature, $T_{meas}$,   to
account for the observed value of F$_{\nu}$(850 ${\mu}$m)/F$_{\nu}$(100 ${\mu}$m).  The values of $R_{det}$ are much larger than the projected radius of the JCMT beam, $R_{JCMT}$, for observations at 850 ${\mu}$m. Therefore, presuming that the JCMT observations have missed most of the submillimeter fluxes from the objects, we estimate the mass of the dust in such a hypothetical detached shell, $M_{dust,det}$ by the expression $M_{dust,det}$ = $M_{dust}$ $(100/T_{meas})(R_{det}/R_{JCMT})^{2}$ where $M_{dust}$ is derived from equation (5) and we assume the dust emits
at temperature $T_{meas}$ instead of 100 K.  
The total mass of  gas
in the detached shell, $M_{det}$, is probably 100 $M_{dust,det}$, and the results for $M_{det}$
are shown in Table 1. The inferred masses in the hypothetical detached shell for SS Lep and 3 Pup are  near 2 M$_{\odot}$  and 80 M$_{\odot}$, respectively.  
If the detached shell is composed of atomic hydrogen, then we would expect a broad
absorption line at Lyman ${\alpha}$ since the line photons that are ``scattered" in the damping portion of the line profile ultimately are absorbed by dust grains.  In fact, for SS Lep, Blondel et al. (1993) report emission at Lyman ${\alpha}$ within 2 {\AA} of line center.
With  ${\tau}$(2 {\AA}) = 1.1 ${\times}$ 10$^{-20}$ $N(H)$ (Diplas \& Savage 1994) where $N(H)_{det}$ (cm$^{-2}$) is the gas column density through the detached shell or
$N(H)_{det}$ = $M_{det}/(4{\pi}m_{H}R_{det}^{2})$, where $m_{H}$ is the mass of a hydrogen atom, then with an estimate that ${\tau}(2 {\AA})$ $<$ 1,  we find $M_{det}$  $<$ 0.30 M$_{\odot}$, nearly a factor of 10 smaller than required.     The absence of CO emission toward SS Lep (Jura \& Kahane 1999) argues against much H$_{2}$ begin present.
It seems unlikely that there
are sufficiently massive detached shells around SS Lep and 3 Pup to explain
the strong submillimeter continuum emission.   

\section{WIND FROM THE DISK?}

Above, we have suggested that the fluxes at ${\lambda}$ $>$ 100 ${\mu}$m are produced by large particles in an orbiting disk. Here, we suggest that the fluxes produced at ${\lambda}$ ${\leq}$ 100 ${\mu}$m may be produced by winds from these disks.    At least  for SS Lep, the M-type companion  probably is 
too warm and its luminosity is too low to have a  stellar wind with enough dust to
account for the infrared fluxes detected by IRAS.      
The red giant in SS Lep has spectral type M4 (Welty \& Wade 1995) while in symbiotic systems with large amounts of dust, the cool mass-losing red giant
stars usually are M6 or later (Kenyon, Fernandez-Castro \& Stencel 1988, Murset \& Schmid 1999).  Also, in Figure 2, we display $F_{\nu}$(12 ${\mu}$m)/F$_{\nu}$(2.2 ${\mu}$m) vs. $M_{K}$ for all the red giants in the
Bright Star Catalog with (B-V) ${\geq}$ 1.5 mag for which distances are
measured with ${\it Hipparcos}$ and for which the infrared fluxes are
reported in the standard catalogs listed in the Figure Caption.  Most of the stars display a value for this ratio characteristic of their photospheric emission.  The stars with a 12 ${\mu}$m
excess all have $M_{K}$ $<$ -6.5 mag while  SS Lep has $M(K)$ = -5.9. Furthermore,  even the stars with $M_{K}$ $<$ -6.5 mag have values of $F_{\nu}$(12 ${\mu}$m)/F$_{\nu}$(2.2 ${\mu}$m) less than that for SS Lep.

A source of the wind could be grain-grain collisions by the large particles in the disk which may  produce small particles which are  driven
out of the system by radiation pressure as they absorb and reprocess the light from the central star.  In this picture, the mass loss rate in a disk wind, ${\dot M}_{wind}$, can be written as (Jura et al. 2000):
\begin{equation}
{\dot M}_{wind}\;=\;(L_{IR}/L_{*})\;[(2\,{\pi}\,R_{in}L_{*})/(c{\overline{\chi}})]^{1/2}
\end{equation}
where $(L_{IR}/L_{*})$ represents the fractional infrared excess radiated by the stellar wind
and ${\overline{\chi}}$ is the average optical opacity of the small grains
in the wind.  For all three stars in Table 1, if we assume that the IRAS fluxes are produced by dust in the wind, then $(L_{IR}/L_{*})$ ${\sim}$ 0.1.  We adopt $R_{in}$ = 3 ${\times}$ 10$^{14}$ cm for SS Lep,  and assume that  $R_{in}$ scales as  $L_{*}^{1/2}$ for the other two stars.
Finally, ${\overline{\chi}}$ ${\approx}$ 2.5 ${\times}$ 10$^{4}$ cm$^{2}$ g$^{-1}$ (Jura et al. 2000).  With these parameters,  values of ${\dot M}_{wind}$ are given in Table 1; the minimum value is  5 ${\times}$ 10$^{17}$ g s$^{-1}$. The nominal lifetime of the disk, $t_{life}$, equals $M_{disk}/{\dot M}_{wind}$, and, as listed in Table 1,  is
 between 10$^{3}$ and 10$^{4}$ yr.  These are minimum ages because  mass may be stored in particles $\gg$ 0.01 cm  which
are not detected in the submillimeter measurements.  In a disk with a lifetime of 1000 yr and a radius of 3 ${\times}$ 10$^{14}$ cm, particles as large as 0.01 cm in radius can grow by coagulation (see Jura et al. 2000). Evidence  for grain evolution around SS Lep is
that its 10 ${\mu}$m silicate feature is unusually broad and
may indicate the presence of crystalline silicates  (Fajardo-Acosta \& Knacke 1995), a possible signature of a long-lived disk (Molster et al. 1999).

\section{CONCLUSIONS}

We have detected mm and sub-mm continuum emission from  two evolved binaries, SS Lep and 3 Pup, and also possibly from BM Gem.    

1.  This continuum is probably produced by emission from dust colder than ${\sim}$ 70 K
lying within 6${\arcsec}$ of the star and  can be explained if the dust particles  are  at least
as large as 0.1 mm in radius.    

2.    We propose that there are circumbinary orbiting disks of at least 5 ${\times}$ 10$^{28}$ g and that the ``large" particles have grown by coagulation in this  disk.  These disks may have winds with mass loss rates of ${\sim}$ 5 ${\times}$ 10$^{17}$ g s$^{-1}$ and lifetimes ${\geq}$ 2000 yr.

\acknowledgements  This work has been partly supported by NASA.  We thank the JCMT staff for their help in acquiring the service mode observations.

\newpage
\begin{center}
FIGURE CAPTIONS
\end{center}
Fig. 1.  A plot of the observed spectral energy distribution for SS Lep with the lower-frequency points labeled by their wavelength (${\mu}$m).  The mm and sum-mm
points are from this paper.   We use the IRAS data for the fluxes from
100 ${\mu}$m to 12 ${\mu}$m, the Two Micron Sky Survey for the flux at 2.2 ${\mu}$m, the {\em Hipparcos} photometry for B and V magnitudes and
the TD-1 satellite photometry for the ultraviolet data.   The solid and dashed lines show the  fluxes for the two stars in the system assumed to be black bodies. 
An extinction law that varies as ${\nu}^{+1}$ with $A_{V}$ = 0.4 mag is assumed.
  The dotted line
shows the IRAS and sub-mm (Knapp, Sandell \& Robson 1993) fluxes for VY CMa scaled by a factor of 0.0109.  The dot-dashed line shows the IRAS and sub-mm (Jura \& Turner 1998, Van der Veen et al. 1994) fluxes for the Red Rectangle scaled by a factor
of 0.159.  The errors are smaller than the squares.
\\
\\
Fig 2.  Plot of $M_{K}$ vs. F$_{\nu}$(12 ${\mu}$m)/F$_{\nu}$(2.2 ${\mu}$m) for
stars in the Bright Star Catalog with (B-V) ${\geq}$ 1.50 mag, parallaxes measured to better than 1${\sigma}$ with the ${\it Hipparcos}$ satellite, 2.2 ${\mu}$m fluxes from the Two Micron Sky Survey (Neugebauer \& Leighton 1969) and non color-corrected 12 ${\mu}$m fluxes from the IRAS survey.  Most of the stars exhibit photospheric values of  F$_{\nu}$(12 ${\mu}$m)/F$_{\nu}$(2.2 ${\mu}$m); but some stars with $M_{K}$ $<$ -6.5 display an excess at 12 ${\mu}$m caused by dust emission.  The unique position of SS Lep in this diagram for ${\sim}$500 stars is evident, with its large 12 ${\mu}$m flux perhaps arising from a disk wind.    
\newpage
\begin{center}
{\bf Table 1 -- Stellar Properties, Observations and Model Results}
\begin{tabular}{lrrr}
\hline
\hline
Property & SS Lep & 3 Pup & BM Gem \\
\hline
$D_{*}$(kpc) & 0.33$^{a}$ & 2.1$^{b}$ & 1.5$^{c}$  \\
$T_{*}$ (K) (component A) & 9000$^{d}$ & 9000 & 3000 \\
$T_{*}$ (K) (component B) & 3500$^{e}$ &   ?    &  ? \\
$R_{*}$ (10$^{12}$ cm) (component A)& 1.1 & 9.5 & 30 \\
$R_{*}$ (10$^{12}$ cm) (component B)& 7.4 & ?&  ? \\
$L_{*}$ (10$^{3}$ L$_{\odot}$) (component A) & 1.5 & 110 & 13 \\
$L_{*}$ (10$^{3}$ L$_{\odot}$) (component B) & 1.5 & ? & ? \\
F$_{\nu}$(100 ${\mu}$m)$^{f}$ (Jy) & 3.8 & 7.3 & 1.3\\
F$_{\nu}$(450 ${\mu}$m)$^{g}$ (mJy) & 148 (23) & 189 (48) & 24 (13) \\
F$_{\nu}$(850 ${\mu}$m)$^{g}$ (mJy) & 58.6 (2.1) & 81.4 (3.3) & 6.6 (1.9) \\
F$_{\nu}$(1350 ${\mu}$m)$^{g}$ (mJy) & 25.4 (2.4) & 46.2 (3.3) & 5 (2) \\
$T_{pred}^{h}$ (K), $p$ = 1 &  123 & ${\geq}$130 & ${\geq}$79 \\
$T_{meas}^{i}$ (K), $p$ = 1&  ${\leq}$ 38  & ${\leq}$ 44 & ${\leq}$ 67 \\
$T_{pred}^{h}$ (K), $p$ = 0 &  46 & 45 & 32 \\
$T_{meas}^{i}$ (K), $p$ = 0 & ${\leq}$600 & ${\infty}$ & ${\infty}$ \\
$M_{dust}^{j}$ (10$^{28}$ g) & 5 & 300 & 10\\
$R_{det}^{k}$ (10$^{17}$ cm) & 5.6 & 28 & 2.0 \\
$M_{det}^{l}$ (M$_{\odot}$) & 2.3 & 77 & 0.01 \\
${\dot M}_{wind}^{m}$ (10$^{17}$ g s$^{-1}$) & 5 & 75 & 15 \\
$t_{life}$ (1000 yr) & ${\geq}$3 & ${\geq}$13& ${\geq}$2\\
\hline
\end{tabular}
\end{center}
$^{a}$ from ${\it Hipparcos}$; $^{b}$ Plets et al. (1995); $^{c}$Kahane et al. (1998); $^{d}$Blondel et al. (1993); $^{e}$ extrapolated from Dyck, Van Belle \& Thompson (1998); $^{f}$measurement from IRAS; $^{g}$fluxes from this work; 
the numbers in parenthesis are the 1${\sigma}$ rms noise errors for each  measurement and do not include uncertainties in the calibration (see text). $^{h}$predicted temperature at 6${\arcsec}$ from the system from equation (2)  and the stellar parameters given in this Table; $^{i}$inferred temperature from equation (3)  and
F$_{\nu}$(850 ${\mu}$m)/F$_{\nu}$(100 ${\mu}$m) from above ; $^{j}$Dust mass from equation (5) with the assumption that $p$ = 0 and
using the observed flux at 850 ${\mu}$m; $^{k}$radius of a detached shell where the grain temperature is expected to be as low as $T_{meas}$ for $p$ = 1; $^{l}$inferred mass of a detached shell; $^{m}$from equation (6)
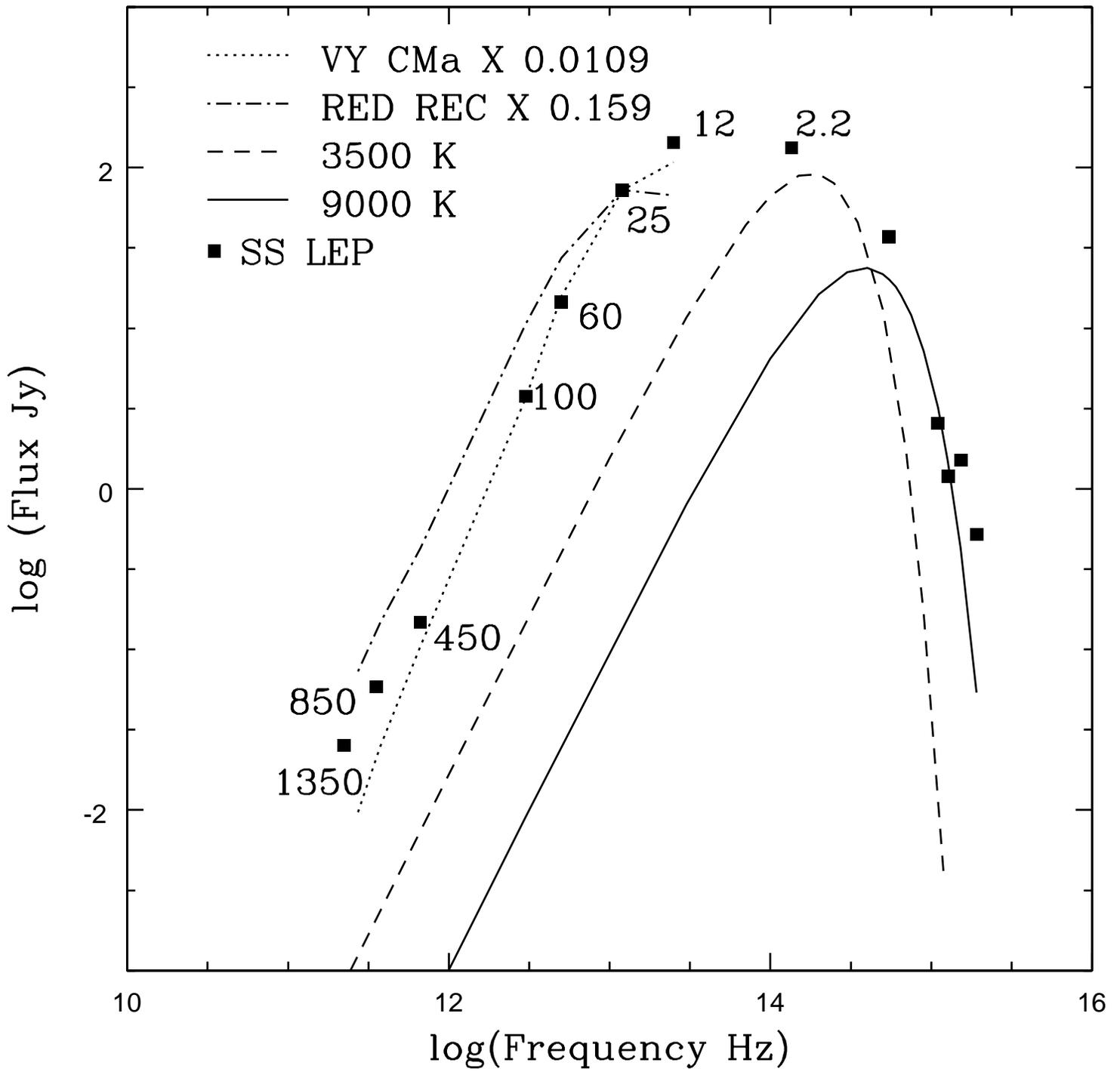
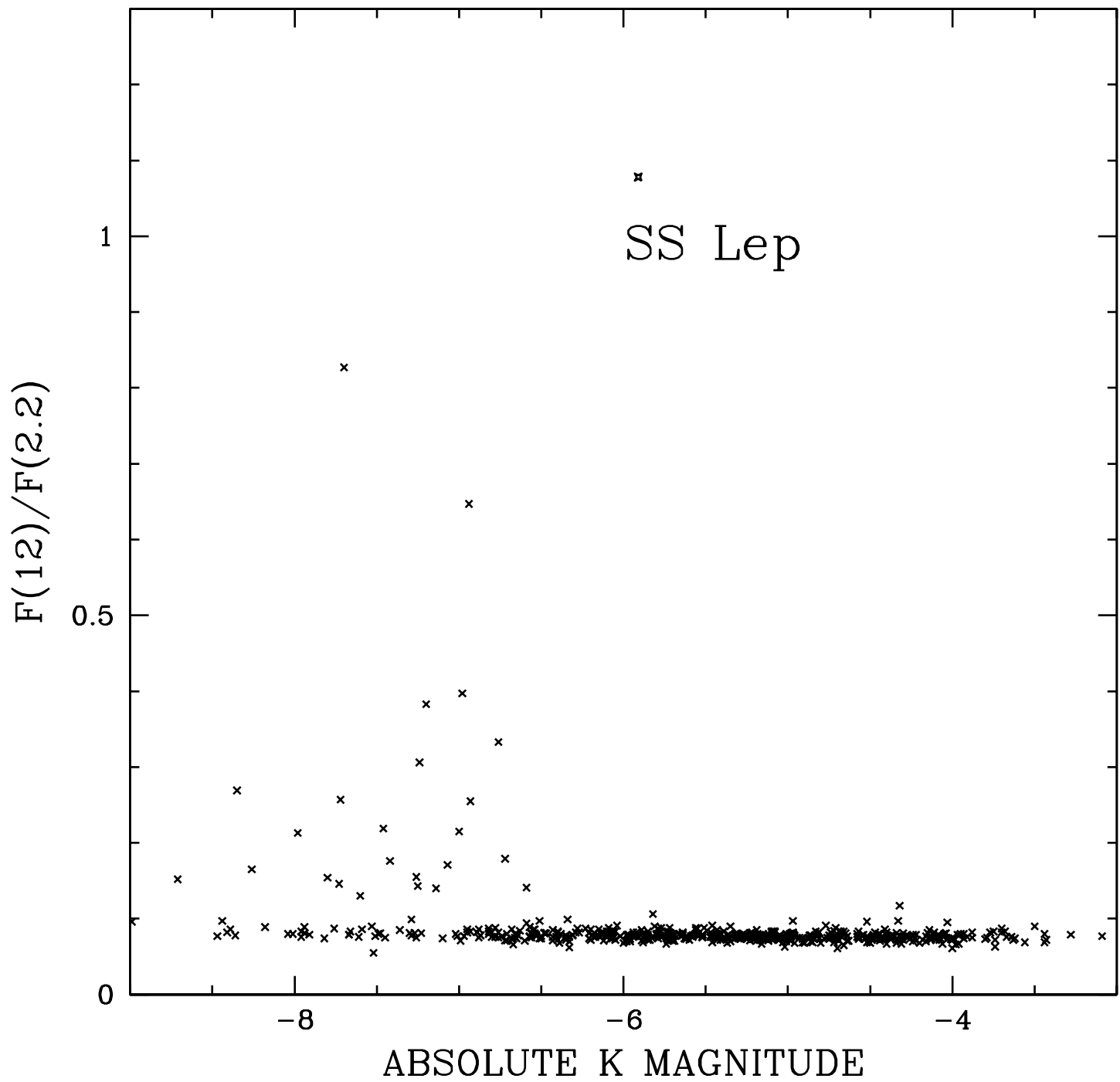
\end{document}